\documentclass[aps,prl,showpacs,twocolumn]{revtex4-1}

\usepackage{amsmath,graphics}
\usepackage[next]{inputenc}
\usepackage[dvips]{epsfig}
\usepackage[colorlinks=true,citecolor=blue,linkcolor=blue]{hyperref}
\usepackage{bbm}
\usepackage{bbold}

\def\be{\begin{equation}}
\def\ee{\end{equation}}

\def\bi{\begin{itemize}}
\def\ei{\end{itemize}}
\def\bn{\begin{enumerate}}
\def\en{\end{enumerate}}
\def\bea{\begin{eqnarray}}
\def\eea{\end{eqnarray}}

\def\ba{\begin{array}}
\def\ea{\end{array}}
\def\bd{\begin{displaymath}}
\def\ed{\end{displaymath}}

\begin{document}
\title{Topological Crystalline Insulators in Transition Metal Oxides}
\author{Mehdi Kargarian}
\author{Gregory A. Fiete}
\affiliation{Department of Physics, The University of Texas at Austin, Austin, TX 78712, USA}

\begin{abstract}
Topological crystalline insulators (TCI) possess electronic states protected by crystal symmetries, rather than time-reversal symmetry. We show that the transition metal oxides with heavy transition metals are able to support nontrivial band topology resulting from mirror symmetry of the lattice. As an example, we consider pyrochlore oxides of the form A$_2$M$_2$O$_7$.  As a function of spin-orbit coupling strength, we find two $\mathrm{Z}_2$ topological insulator phases can be distinguished from each other by their  mirror Chern numbers, indicating a different TCI. We also derive an effective $\bf{k\cdot p}$ Hamiltonian, similar to the model introduced for $\mathrm{Pb_{1-x}Sn_{x}Te}$, and discuss the effect of an on-site Hubbard interaction on the topological crystalline insulator phase using slave-rotor mean-field theory, which predicts new classes of topological quantum spin liquids.      
\end{abstract}
\date{\today}
\pacs{71.10.Fd,71.70.Ej,75.10.Kt} 
\maketitle
\textit{Introduction}.$-$A large class of materials has been predicted to possesses a nontrivial $\mathrm{Z}_2$ topological classification \cite{Bernevig:sci06,Zhang:np09,Hasan:rmp10,Qi:rmp11,HasanMoore:Ann11}. These materials have gapless surface modes protected by time-reversal symmetry. However, crystal symmetries can also impose topological features, which leads to an additional source of topological protection in materials possessing these symmetries \cite{Fu:prb07,Teo:prb08,mong:prb10,Li:np10,Slager:nap13}. One class of such insulators is topological crystalline insulators (TCI) with surface states protected by either point \cite{Fu:prl11} or mirror symmetry \cite{Hsieh:naC2012}. The recent discovery of TCI in the narrow band semiconductor $\mathrm{Pb_{1-x}Sn_{x}Te/Se}$ has paved the way to a new class of topological materials beyond the $\mathrm{Z}_2$ topological insulators (TI) \cite{Xu:naC2012,Tanaka:np2012,Dziawa:nm12}. However, it is important to extend the domain of candidate TCI to include other materials, such as those with spontaneously broken time reversal symmetry (TRS) through magnetic ordering resulting from electron interactions. 

\begin{figure}[hb]
\begin{center}
\includegraphics[width=8cm]{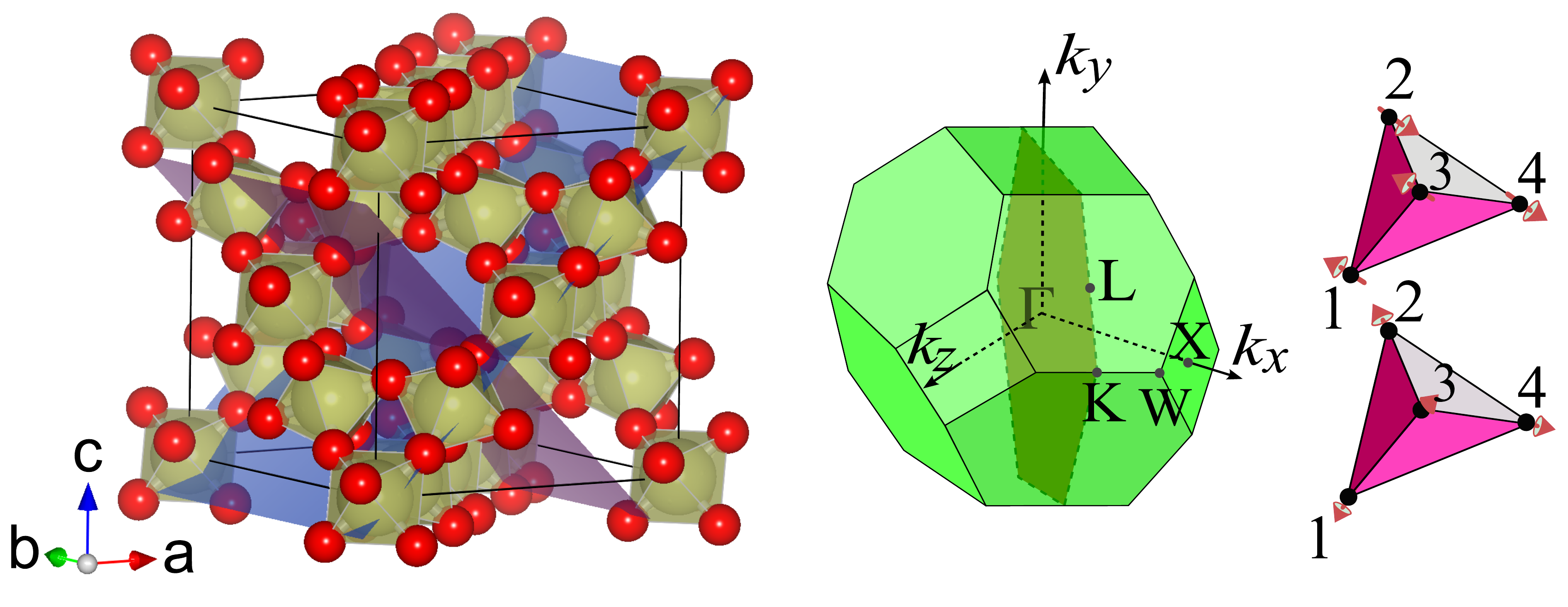}
\caption{(Color online) Left: Pyrochlore lattice. Transition metal ions (large balls) are located on the corner shared tetrahedra, each surrounded by an oxygen cage (red balls). Rare-earth elements fill spaces between cages. Two mirror planes are also shown. Center: Brillouin zone of face centered cubic lattice with mirror plane. Capital letters denote the high symmetry points. Right: a representation of unit cell with four sites and two possible magnetic orderings. On top ordering preserve mirror symmetry and on bottom ordering is of all-in-all-out which breaks mirror symmetry.} \label{lattice}
\end{center}
\end{figure} 

In this Letter, we show that transition metal oxides (TMO) with 5$d$ orbitals are candidate materials for a TCI, extending the list of potential TCI materials to include those where interaction effects could play a role in driving qualitatively new phases and phenomena. Pyrochlore iridates, such as $\mathrm{A_2Ir_2O_7}$, where A is a rare-earth element, have been shown to possess a nontrivial $\mathrm{Z}_2$ invariant \cite{Pesin:np10,Kargarian:prb11,Yang:prb10b} so long as the strength of direct $d$-$d$ hopping relative to indirect hopping via the oxygen orbitals does not fall within a certain window \cite{Witczak:prb12} and time-reversal symmetry is not broken \cite{Go:prl12}.  We show that surface states exist even if TRS is broken by the application of a magnetic field or from local magnetic moments induced by interactions. We argue that these surface sates are protected by mirror symmetry, and originate from a nonzero mirror Chern number \cite{Teo:prb08,Hsieh:naC2012}. This nontrivial band topology leads to a new topological Mott phase in the regime of intermediate Hubbard interaction where the charge degrees of freedom on the surface become gapped, but the TCI structure of the gapless surface spin modes remain intact (a type of topological quantum spin liquid): the topological crystalline Mott insulator (TCMI).    
     
\textit{Model}.$-$The transition metal oxide $\mathrm{A_2Ir_2O_7}$ has a pyrochlore lattice structure (see Fig.\ref{lattice}) composed of corner shared tetrahedra of $\mathrm{Ir^{+4}}$ ions ($d$-shell filling f=1/2).  Each $\mathrm{Ir^{+4}}$ is surrounded by an octahedral cage of oxygen which splits the atomic $d$-levels into a lower $t_{2g}$ and higher $e_g$ manifold; spin-orbit coupling further acts within the $t_{2g}$ states to create eigenstates of total angular momentum.  A minimal tight-binding model describing hopping of electrons between transition ions is,
\bea
 \label{Hd} 
 H_{d}=\sum_{i}t^{\gamma\gamma'}_{i}d^{\dag}_{i\gamma}d_{i\gamma'}+\sum_{<ij>}(T^{\gamma\gamma'}_{o,ij}+T^{\gamma\gamma'}_{d,ij})d^{\dag}_{i\gamma}d_{j\gamma'},
 \eea 
 where first term contains both onsite energy $\varepsilon_d$ and local spin-orbit coupling with strength $\lambda$ as $t_i=\varepsilon_d-\lambda \bf{l}\cdot \bf{s}$, and the second term describes nearest-neighbor hopping between transition metal ions \cite{Pesin:np10,Kargarian:prb11,Witczak:prb12}. Here, $\gamma$ is a collective index including both $t_{2g}(yz,zx,xy)$ orbitals and spins, $T_{o,ij}$, $T_{d,ij}$ are matrices for the oxygen mediated and direct hopping integrals, respectively.  For simplicity, we consider only oxygen mediated hopping of electrons which captures the correct physical picture over a wide range of direct $d$ hopping values \cite{Witczak:prb12}. Outside of this range, the system is gapless and a Weyl semi-metal can result from interaction effects \cite{Witczak:prb12,Wan:prb11}. Physical pressure may provide a route to tune between these two interesting regimes.
 
Because of the cubic structure of the lattice (face centered cubic with four point basis), the lattice is symmetric with respect to a variety of mirror planes; two are shown in Fig.\ref{lattice}. Crystal surfaces such as (010) are symmetric about the mirror planes. We show they support topologically protected gapless surface modes. The mirror operator transforming orbitals and spin about the mirror plane ($\bar{1}$01) is $M=R\otimes U$, where $R$ transforms the orbitals in local coordinates of each rotated octahedra and $U$ rotates spin by $\mathrm{180^{o}}$ about an axis normal to the mirror plane, 
\bea
 R=\left(
  \begin{array}{cccc}
  r_1&o&o&o\\ 
 o&o&o&r_2\\ 
 o&o&r_3&o\\
 o&r_4&o&o\\ 
  \end{array}\right),~~U=\frac{i}{\sqrt{2}}\left(
  \begin{array}{cc}
  -1&1\\ 
 1&1\\  
  \end{array}\right),
\eea where $r_i$ is a reflection matrix transforming local $t_{2g}$ orbitals on sites $i$=1..4 (see Fig.\ref{lattice}), and $o$ is a 3$\times$3 zero matrix. The $r_i$ are given by
\bea \label{r_martices} r_1=r_2=r_4=\left(
  \begin{array}{ccc}
  1&0&0\\ 
 0&0&-1\\ 
 0&-1&0\\ 
  \end{array}\right),~
  r_3=\left(
  \begin{array}{ccc}
  0&1&0\\ 
 1&0&0\\ 
 0&0&1\\ 
  \end{array}\right).
   \eea 
 The transformation of spin under the matrix $U$ is used to determine which pattern of magnetic orderings or magnetic field preserve the mirror symmetry about the ($\bar{1}$01) plane. Note reflection about this mirror plane takes a unit cell located at lattice vector $\textbf{R}_{i}=(x_i,y_i,z_i)$ to $\bar{\textbf{R}}_{i}=(z_i,y_i,x_i)$.  The k-space Hamiltonian transforms as $MH_d(k_x,k_y,k_z)M^{-1}=H_d(k_z,k_y,k_x)$. On the mirror plane in the Brillouin zone (see Fig.\ref{lattice}) where $k_x=k_z$, [$H_d,M$]=0.  Therefore, on this plane the Bloch eigenstates of $H_d$ can be labeled by mirror eigenvalues $\pm i$ since $M^2=-1$ \cite{Teo:prb08,Hsieh:naC2012}.
  
\textit{Mirror Chern number}.$-$ For 0$<$$\lambda/t$$<$2.8 the non-interacting model is metallic; larger values of $\lambda$ open a gap in the spectrum and result in a $\mathrm{Z}_2$ TI with indicies $(1;000)$ \cite{Pesin:np10}.  This gap closes for $\lambda_c$$\approx$3.25$t$ by forming nodes at the four equivalent L points in the Brillouin zone. These nodes will again be gapped for $\lambda$$>$$\lambda_c$, but inverted relative to $\lambda$$<$$\lambda_c$.  However, the $\mathrm{Z}_2$ index remains unchanged because an even number of band inversions occurred. We find the following parity invariants, $\delta(\Gamma_a)$ in the notation $(\delta,\Gamma_a)$ for $\lambda/t=3$: $(+1,\Gamma), (-1,L),(-1,X)$ for $\lambda/t=5$: $(+1,\Gamma), (+1,L),(-1,X)$.   
The gap closing phenomenon persists in the presence of interactions \cite{Pesin:np10}.  We show these seemingly identical TI phases can be distinguished by a mirror Chern number. A similar gap closing at the L points also occurs in $\mathrm{Pb_{1-x}Sn_{x}Te}$, separating a trivial insulator from a TCI \cite{Hsieh:naC2012}.   

\begin{figure}[t]
\begin{center}
\includegraphics[width=9cm]{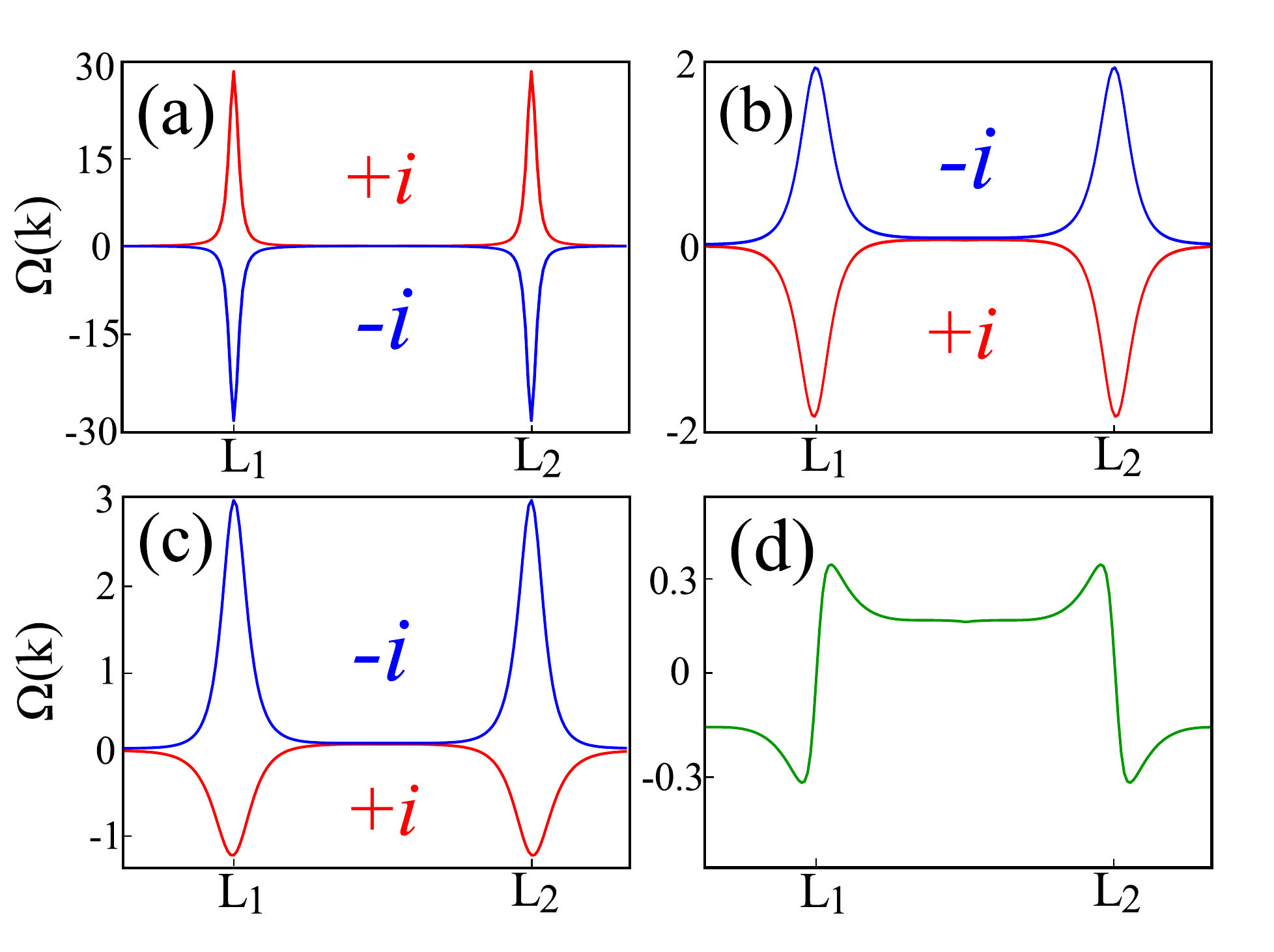}
\caption{(Color online) Berry curvature $\Omega(\textbf{k})$ along the boundary of the mirror plane in Brillouin zone. $L_{1}=(\pi,\pi,\pi)$ and $L_{2}=(\pi,-\pi,\pi)$ are high symmetry points of 3D Brillouin zone residing on the boundary of the mirror plane in k-space. In (a)-(b) the time reversal symmetry is preserved and $\lambda$$<$$\lambda_c$ in (a) and $\lambda$$>$$\lambda_c$ in (b). In (c)-(d) the time reversal symmetry is broken by the magnetic ordering shown in Fig.\ref{lattice} which preserves and breaks the mirror symmetry in (c) and (d), respectively.} \label{Berry}
\end{center}
\end{figure}

The fact that all bands on the mirror plane can be labeled by mirror eigenvalue $\pm i$ means that each band is effectively spin polarized. We have calculated the average values of the spin operator $<$$\textbf{S}$$>$ for each band and found for mirror related bands $<$$\textbf{S}$$>\propto$$\pm\frac{S}{\sqrt{2}}(\hat{x}-\hat{z})$, which is perpendicular to the mirror plane. We also calculated the Berry curvature $\Omega(\textbf{k})=\nabla\times \textbf{A}$, where $\textbf{A}=i\sum_{n}\langle u_n(\textbf{k})|\nabla|u_n(\textbf{k})\rangle$ is the Berry connection (summed over all occupied bands) on the mirror plane in the Brillouin zone (see central panel in Fig.\ref{lattice}). In Fig.\ref{Berry}a-b we plot $\Omega^{\pm i}(\textbf{k})$ for occupied bands with mirror eigenvalues $\pm i$ for $\lambda$$<$$\lambda_c$ and $\lambda$$>$$\lambda_c$. The path in k-space is chosen along the mirror Brillouin zone boundary including $L_1=(\pi,\pi,\pi)$ and $L_2=(\pi,-\pi,\pi)$. The main contribution to $\Omega(\textbf{k})$ comes from k points around L. The plots clearly reveal that $\Omega^{+i}(\textbf{k})-\Omega^{-i}(\textbf{k})$ changes sign upon gap closing. For each polarization we calculated the Chern number \cite{Teo:prb08}, $n_{M}=(n_{+i}-n_{-i})/2$, and found that for $\lambda<\lambda_c$: $n_{+i}$=+1 and $n_{-i}$=-1 yielding $n_{M}$=+1 and for $\lambda>\lambda_c$: $n_{M}$=-1. 
The nonzero value of mirror Chern number shows that the mirror symmetry gives rise to nontrivial band topology and it proves that the $\mathrm{Z}_2$ TI phases around the gap closing point can be further labeled by $n_{M}$=$\pm 1$. 

This behavior persists even in the presence of time reversal breaking perturbations, which may result from either a magnetic field or from local magnetic moments. As an example, we add $H'=\sum_{i}\textbf{B}_{i}\cdot\textbf{S}_i$ to Eq.\eqref{Hd}. In order to understand which patterns of local moments or magnetic field preserve the mirror symmetry, we note that under reflection about the mirror plane: 
$\sigma^{x}\rightarrow - \sigma^{z},
\sigma^{y}\rightarrow -\sigma^{y},
\sigma^{z}\rightarrow -\sigma^{x}$.  
Hence, $H'$ is invariant if $\textbf{B}_{i}$=$\pm$B(-1,0,1), or any magnetic ordering which is perpendicular to mirror plane. Two configurations of magnetic orderings are shown in Fig.\ref{lattice}. The ordering on the top has mirror symmetry, while the all-in-all-out configuration breaks the mirror symmetry. Such orderings can be stabilized by interactions \cite{Wan:prb11,Witczak:prb12} 

Consider a configuration of moments that preserve the mirror symmetry. The mirror eigenvalues are still well defined and can be used to label the Bloch states. In Fig.\ref{Berry}c we plot the corresponding Berry curvatures. Although these values of Berry curvature look asymmetric between states with mirror eigenvalues $+i$ and $-i$, integrated over the entire mirror Brillouin zone they are the same up to a minus sign, namely $n_{+i}=-n_{-i}$ yielding $n_{M}$=-1. This shows that the band topology is constrained by mirror symmetry, not by TRS.  If the mirror symmetry is broken, say by an all-in-all-out ordering, the Bloch states are no longer eigenstates of the mirror operator. The corresponding Berry curvature is shown in Fig.\ref{Berry}d. Due to its antisymmetric nature around each L point, the Berry phase of each L point and thus the total Chern number is zero. While the two magnetic orderings yield zero total Chern number, the mirror symmetric one results in an integer value of the mirror Chern number.
 
\begin{figure}[t]
\begin{center}
\includegraphics[width=8cm]{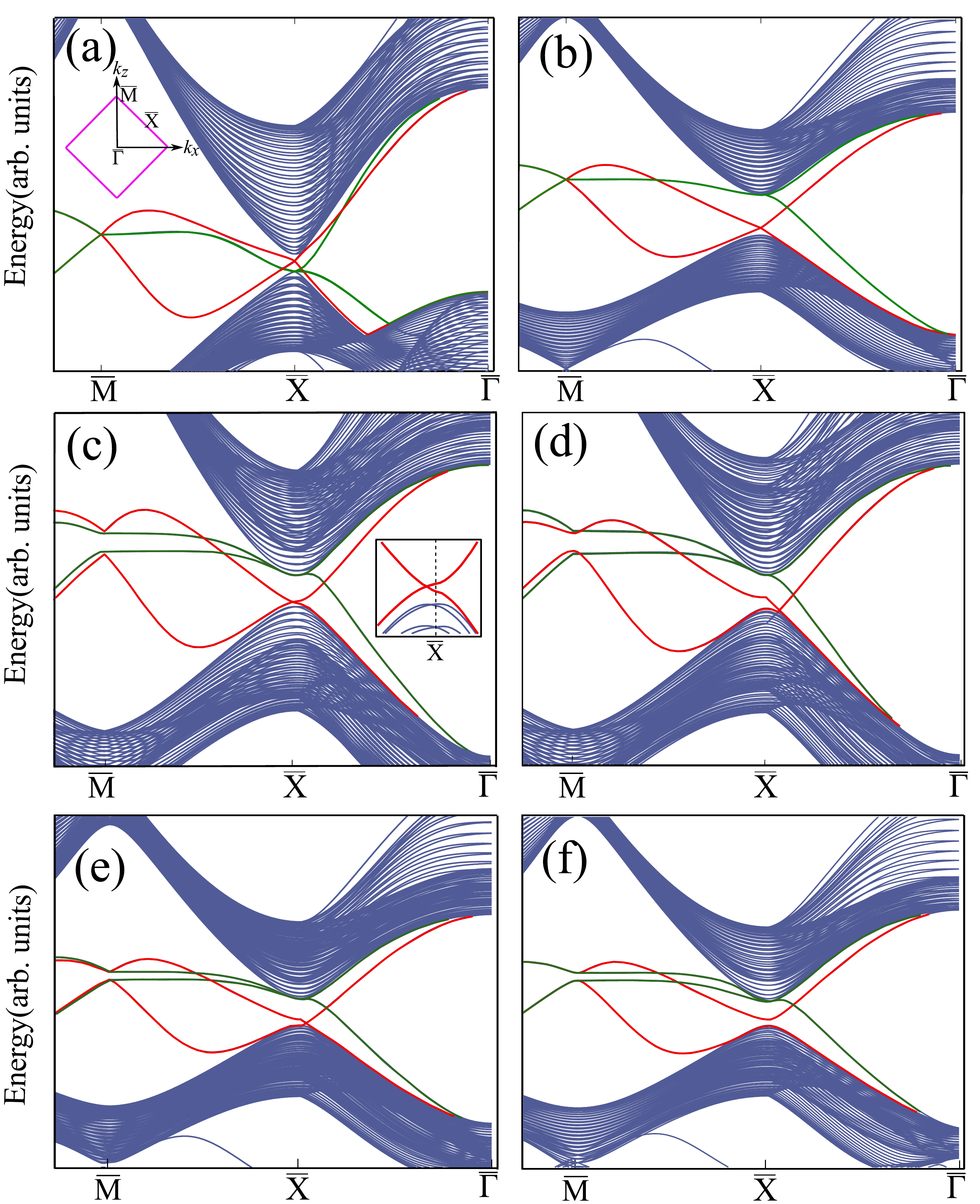}
\caption{(Color online) The band structure of the tight-binding model in Eq.\eqref{Hd}  along the high symmetry points of projected Brillouin zone as shown in inset of (a).  In panels (a)-(b) the time reversal symmetry is preserved and $\lambda$=3 (a) $\lambda$=4 (b). In panels (c)-(e) the time reversal symmetry is broken, but the mirror symmetry is preserved: (c) magnetic field \textbf{B}=B(1,0,1), inset enlarges the crossing (d) magnetic field \textbf{B}=B(-1,0,1) and (e) magnetic ordering. In panel (f) both time reversal and mirror symmetries are broken by a magnetic ordering of all-in-all-out type. Green and red lines correspond to surface sates on bottom and top surfaces, respectively, of slab. Dark blue lines are bulk sates.} \label{surfacestates}
\end{center}
\end{figure}

\textit{Surface states}.$-$The bulk-boundary correspondence implies the surface between two insulators with different bulk band topology carries gapless surface modes \cite{Hasan:rmp10,Qi:rmp11}. In order to confirm that the surface of the pyrochlore lattice supports gapless surface states we consider a slab geometry along the [010] direction. This slab, as shown in Fig.\ref{lattice}, is symmetric about two mirror planes: ($\bar{1}$01) and (101) with an offset of 1/4 (or 3/4) of lattice spacing. We diagonalize the Hamiltonian in this geometry and plot the surface states along the high symmetry points of the projected Brillouin zone. The results for different sets of parameters are shown in Fig.\ref{surfacestates}. For a time reversal symmetric system there are Dirac nodes right at the projected time reversal invariant momenta (TRIM) $\bar{M}$ and $\bar{X}$ (see Fig.\ref{surfacestates}a-b). By increasing spin-orbit coupling across the gap closing point at $\lambda_c$ the overall features of the surface states remain unchanged except that the area of Fermi surface enclosed by Dirac nodes at $\bar{M}$ change and its electron or hole like character on the two surfaces of the slab change. The degeneracy of the green surface modes along $\bar{M}$$\bar{X}$ can be lifted by applying different onsite potentials on sites 1 and 3 (see unit cell in Fig.\ref{lattice}).        

Now we break the time reversal symmetry by adding $H'$. We first consider the cases in which the perturbation respects the mirror symmetry. The results for uniform magnetic field $\textbf{B}_1$=B(1,0,1) and $\textbf{B}_2$=B(-1,0,1) are shown in Fig.\ref{surfacestates}(c)-(d), respectively. Note $\textbf{B}_1$ breaks the symmetry about mirror plane ($\bar{1}$01) but preserves the mirror symmetry about plane (101), while $\textbf{B}_2$ does the opposite. The Kramers degeneracy is lifted and Dirac nodes become gapped at $\bar{M}$ and $\bar{X}$ point. However, adjacent to $\bar{X}$ there are still some crossing surface modes (see inset in Fig.\ref{surfacestates}(c)). Mirror planes in Brillouin zone, such as those shown in Fig.\ref{lattice}, define mirror lines in the projected Brillouin zone. For example, for $k$ points on the line $\bar{\Gamma}\bar{X}$, where $k_x$=$k_z$, the eigenstates can be labeled by mirror eigenvalues. The same arguments hold for line $\bar{M}\bar{X}$. Thus, surface states can also be labeled by their corresponding mirror eigenvalues. The crossing of the surface states along the $\bar{M}\bar{X}$ line is protected by mirror symmetry about (101) plane (see Fig.\ref{surfacestates}c), and the crossing along $\bar{\Gamma}\bar{X}$ line is protected by mirror symmetry about ($\bar{1}$01) plane (see Fig.\ref{surfacestates}d). Away from these lines the degeneracy is lifted. We also calculated the surface states in the presence of a pattern of magnetic ordering which preserves symmetry about the mirror plane ($\bar{1}$01). The result is shown in Fig.\ref{surfacestates}e. The surface modes cross each other along $\bar{\Gamma}\bar{X}$ line as expected. This identification clearly shows that the crossing surface modes are associated with mirror symmetry, independent of the details of a time reversal breaking perturbation.  

Finally, we break symmetry with respect to both mirror planes by considering an all-in-all-out magnetic configuration. The band structure is shown in Fig.\ref{surfacestates}f. It is seen that the degeneracy is lifted and there is no crossing along either $\bar{M}\bar{X}$ or  $\bar{\Gamma}\bar{X}$. We also explored many other cases (not shown here). All mirror symmetry breaking perturbations remove the band crossing in the gap. This finding further verifies that the surface crossing discussed in the previous paragraphs are associated with mirror symmetry. 

\textit{{\bf k$\cdot$p} Hamiltonian}.$-$ It is instructive to give a ${\bf k\cdot p}$ expansion of Eq.\eqref{Hd} near the $L$ point. We first rotate $(k_x,k_y,k_z)$ into a new orthogonal basis $(k_1,k_2,k_3)$ in which $k_1$ is along the $\Gamma L$ line and $k_3$ is perpendicular to mirror plane. Thus, the mirror plane corresponds to $k_3=0$. Right at the $L$ point Eq.\eqref{Hd} in k-space becomes block diagonal with Bloch eigenstates localized on either site 1 (or sites 234) [see unit cell in Fig.\ref{lattice}]. This effectively subdivides the lattice sites in two sets and defines a basis to expand Eq.\eqref{Hd} around the $L$ point. We label them as $\psi_A$ and $\psi_B$, respectively. With this identification the effective Hamiltonian near the $L$ point is,
\bea 
\label{kpH} 
H^{\pm i}=m\tau^{z}\pm\textbf{v}_1\cdot \textbf{k} \tau^{x}+\textbf{v}_2\cdot \textbf{k} \tau^{y},
\eea
where $\tau^{z}=\pm 1$ corresponds to $\psi_A$ and $\psi_B$, respectively, and $\textbf{k}=(k_1,k_2)$. Eq.\eqref{kpH}, and therefore its underlying physics, is not dissimilar from the model introduced to describe the insulator phases of the semiconductor $\mathrm{Pb_{1-x}Sn_{x}Te}$ upon doping \cite{Hsieh:naC2012}. In fact, in both models the character of conduction/valence bands-- $d$-orbital $\psi_A/\psi_B$ in our model and $p$-orbital cation/anion in the $\mathrm{Pb_{1-x}Sn_{x}Te}$ model--gets switched at $L$ point by tuning spin-orbit coupling $\lambda$ and Pb doping, respectively. The band inversion changes the sign of the mass term $m$ in Eq.\eqref{kpH}, which in turn changes the Chern number for $\pm i$ mirror eigenstates by $\pm 1$. Since the gap closing occurs at two equivalent $L$ points (related by $\mathrm{180^{o}}$ rotation about [101] axis) on the mirror plane, the total change of Chern number for each $\pm i$ states will be $\pm 2$. This change of Chern number is consistent with Berry curvatures calculated in Fig.\ref{Berry}. There is, however, a significant difference between our model and the model for $\mathrm{Pb_{1-x}Sn_{x}Te}$. In the latter model, the gap closing at $L$ points signals a topological phase transition between a trivial insulator $\mathrm{PbTe}$ and topological insulator $\mathrm{SnTe}$ with mirror Chern number $n_M=2$. But in our $d$-orbital model the topological phase transition occurs between two topological phases with different mirror Chern number: $n_M=-1$ before and $n_M=+1$ after a gap closing. This distinction between topological insulator phases has some implications for the surface states which can potentially be explored in experiment. For example, on the mirror line $\bar{X}\bar{\Gamma}\bar{X}$ there is exactly one crossing surface mode due to $n_M=+1$, while for $\mathrm{Pb_{1-x}Sn_{x}Te}$ there are two crossings.

\textit{Topological crystalline Mott insulator}.$-$In $d$-orbital models the electron correlations can be strong and might even lead to new topological phases, such as the topological Mott insulator (TMI) \cite{Pesin:np10}. The TMI phase can be obtained through a spin-charge separation inherently included in the slave-rotor decomposition \cite{florens:prb04,florens:prb02}: $c_{j\gamma}=e^{i\theta_{j}}f_{j\gamma}$. The mean-field theory can be written as $H=H_{f}+H_{\theta}$, where $H_f$ and $H_{\theta}$ describe, respectively, spinon and rotor parts of the model, and are related to each other via mean-field parameters $Q_{f}=\langle e^{-i(\theta_{i}-\theta_{j})}\rangle$ and $Q_{\theta}=\langle \sum_{\gamma\gamma'}T_{ij}^{\gamma\gamma'}f^{\dag}_{i\gamma\sigma}f_{j\gamma'\sigma'} \rangle$ which are determined self-consistently as a function of Hubbard interaction \cite{Pesin:np10,Kargarian:prb11}. The Hamiltonian of the electrons is described by Eq.\eqref{Hd} with $d_{i\gamma}\rightarrow f_{i\gamma}$ and $T_{ij}\rightarrow Q_{f}T_{ij}$. Thus, the electron correlations systematically renormalize the band width of the system by decreasing $Q_f$. Mott physics occurs when the quasiparticle weight $Z=$$\langle e^{i\theta}\rangle$ vanishes for sufficiently strong Hubbard interactions, where the bosonic rotor excitations are gapped and become uncondensed. Beyond this limit the low energy excitations are described solely by a spinon Hamiltonian, $H_f$, which may have topological band structure. We calculated the mirror Chern number in the boson uncondensed phase and established that $n_M=-1$. This already means that the previous topological crystalline insulator phase in the weak interaction limit will transit into a Mott analogue--the topological crystalline Mott insulator (TCMI)--which is similar in physics to a topological Mott insulator phase inherited from the nontrivial band topology of a noninteracting TI only with gapless spin modes protected by the mirror symmetry rather than TRS.

\textit{Summary}.$-$We discussed the possible realization of topological crystalline insulators in transition metal oxides with $5d$ orbitals and mirror symmetry, and shown electron interaction effects can drive a novel TCMI phase. One may wonder if the TCI phase will survive disorder in real materials; the experimental examples\cite{Xu:naC2012,Tanaka:np2012,Dziawa:nm12} prove that it does, much as the salient properties of the time-reversal invariant Z$_2$ TIs survive in the presence of the Earth's weak magnetic field. Indeed, theory supports the robustness of the TCI phase if mirror symmetry is broken locally \cite{Hsieh:naC2012}.

We thank Liang Fu for helpful discussions and acknowledge financial support through ARO Grant No. W911NF-09-1- 0527, NSF Grant No. DMR-0955778, and by grant W911NF-12-1-0573 from the Army Research Office with funding from the DARPA OLE Program.

%

\end{document}